\documentclass{aa}  
\usepackage{graphicx}
\usepackage{txfonts}
\begin{document}

\title{Equidistant frequency triplets in pulsating stars: The Combination Mode Hypothesis}

\author{M. Breger\inst{1} \and K. Kolenberg\inst{1,2}}
      
\offprints{M. Breger}

\institute{Department of Astronomy, University of Vienna, T\"urkenschanzstrasse 17, A-1180 Vienna\\
              \email{michel.breger@univie.ac.at}
\and 
Institute of Astronomy, University of Louvain,
              Celestijnenlaan 200B, B-3001 Heverlee, Belgium\\
              \email{kolenberg@astro.univie.ac.at}}


\date{Accepted 2005 month day.
      Received 2005 month day;
      in original form 2005 month date}

\abstract{

Multiplet structures are a common feature in pulsating stars, and can be the consequence of rotational splitting,
 mode interaction or sinusoidal amplitude variations.  In this paper we examine the phenomenon of (nearly) equidistant triplets, which
  are unlikely to be caused by rotational splitting, in
  different types of pulsating stars: a $\delta$~Scuti variable (1 Mon), 
  an RR Lyrae variable (RR Lyr) and a short-period Cepheid (V743 Lyr).
  We examine the hypothesis that one of the modes forming the triplet results from a combination of
  the other two modes. 
  The analyses were carried out on recent data sets by using
  multiple-frequency analyses and statistics with the package PERIOD04.
  In particular, the small departures from equidistance were calculated
  for the three selected stars.  For the $\delta$~Scuti variable 1 Mon, the
  departure from equidistance  is only 0.000079 $\pm$ 0.000001 cd$^{-1}$ (or
  0.91 $\pm$ 0.01 nHz). For 1~Mon the Combination Mode Hypothesis with a mode
  excited by resonance is the most probable explanation. For the star RR
  Lyr, the hypothesis of resonance through a combination of modes should be considered.
  The results for the best-studied cepheid with a Blazhko period (V743~Lyr) are inconclusive because of an unfavorable period of 1.49d and insufficient data. }

\titlerunning{Equidistant frequency triplets in pulsating stars}
\maketitle

\section{The problem}

The different frequencies excited in multimode stellar pulsators are an important tool to increase our understanding of
stellar structure and evolution.  Equidistant or near-equidistant frequency triplets form a
special case, and are found in different types of pulsators. Here the frequency triplets
have different physical origins which need to be examined carefully. Examples are white
dwarfs (rotational splitting of nonradial modes: e.g., Winget et al. 1991), sdB stars (see Vu\v{c}kovi\'c et al. 2006 for a puzzling triplet) and roAp stars (oblique pulsator model).
In this paper we examine a $\delta$~Scuti star (1 Mon: see Shobbrook \& Stobie 1974), a RR Lyrae star (RR Lyr: see Smith et al. 2003; Kolenberg et
al. 2006), as well as a cepheid (V743 Lyr: equidistant spacing reported by Koen 2001).

Equidistant or nearly equidistant frequency triplets can be explained in several ways:

(i) {\em Amplitude variability:} Amplitude variability gives rise to a number of separate peaks in a Fourier diagram. The number, height and separation of these peaks depend on the data coverage as well as the type of amplitude variability. A very special case exists: sinusoidal amplitude variability leads to an equidistantly spaced frequency triplet with two side peaks of equal height. This was originally proposed to explain the roAp stars (Kurtz 1982)
through the Oblique Pulsator Model. In general, the peaks are not equidistant: to avoid the common spurious multiple peaks caused by amplitude variability and limited coverage, independent data samples need to be compared to each other.

(ii) {\em Rotational splitting of nonradial modes: physical and geometric effects:} Rotational splitting of nonradial
modes is a promising explanation for equidistant triplets only for stars that rotate extremely slowly (such as white dwarfs).
Depending on which modes are examined, the rotational splitting may correspond to the rotation of the interior,
which provides an important tool for asteroseismolgy. 
Second-order effects in rotational splitting become very important for modes dominant in stellar envelopes. In the presence of stellar rotation in excess of a few km/s, p-mode splitting will not be symmetric and equidistant triplets will not be seen (Pamyatnykh 2000).
An example is the moderately rotating (66 $\pm 16$ km s$^{-1}$) $\delta$ Scuti star FG Vir: an $\ell$ = 1 triplet is observed to
have separations of 0.42 and 0.64 c/d (Zima et al. 2006).

(iii) {\em Successive radial orders:} In the asymptotic pulsation limit, successive radial orders are equidistantly spaced.
If three successive radial orders are excited, an equidistant
frequency triplet can result. Such cases can be recognized easily because of the relatively large and predictable frequency
differences between radial orders and will not be considered here.

(iv) {\em The Combination Mode Hypothesis:} This hypothesis, examined in the next section, requires an exact equidistant frequency triplet.
In the case of resonance excitation, an almost equidistant frequency triplet
also fits the hypothesis.

\section{Combination frequencies as an explanation for triplets}

The occurrence of linear combinations of frequencies is a common feature in
pulsation phenomena.  Such combinations often
occur in the frequency spectra of stars pulsating with multiple oscillation
modes and are even seen in small-amplitude pulsators such as $\delta$~Scuti stars (Breger et al. 1999)
and white dwarfs (Wu 2001).

In this paper we study the possibility of triplet structures resulting
from mode combination and/or resonance phenomenon.
The basic idea of the combination mode hypothesis is illustrated in Fig.\,1.
Assume that two modes f$_1$ and f$_2$ are excited and let f$_1$ have the
highest amplitude. As a consequence, its harmonic frequency $2f_1$ is also
likely to occur in the frequency spectrum.  Since a real second mode is excited at $f_2$,
the combination $2f_1-f_2$ may also occur. This combination is observed at
$f_3$ and as a consequence a triplet is observed around $f_1$. 
The difference $f_2-f_1$ is referred to as $f_B$, the beat frequency
(Figure\,1).  It
plays a crucial role in understanding RR Lyrae stars showing the Blazhko effect (see
Section\,4). 

It is possible that the left-side or lower-frequency peak in the triplet, $f_3$, may not be located at the exact frequency of the
combination. If there exists another mode in the pulsation spectrum of the
star with an eigenfrequency close to that of the combination mode frequency, it may be
excited through a resonance with the combination mode frequency and gain
observable amplitudes. This additional mode has a frequency at or close to
the combination frequency. Resonances are known to play important roles in determining
the pulsational behavior of stars and we refer to an excellent theoretical discussion by
Buchler et al. (1997).

Consequently, one should examine the departure from equidistance
in the triplet:

\begin{equation}
 \delta f = (f_2 - f_1) - (f_1 - f_3) = f_2 + f_3 - 2f_1.
\end{equation} 

If the triplet is not exactly equidistant, i.e., $\delta f \neq 0$, it is possible to
exclude that the frequency is a pure combination frequency. This would 
strengthen the case of a resonance invoked by a combination frequency.

\begin{figure}[htb!]
\centering
\includegraphics[angle=0,width=9cm]{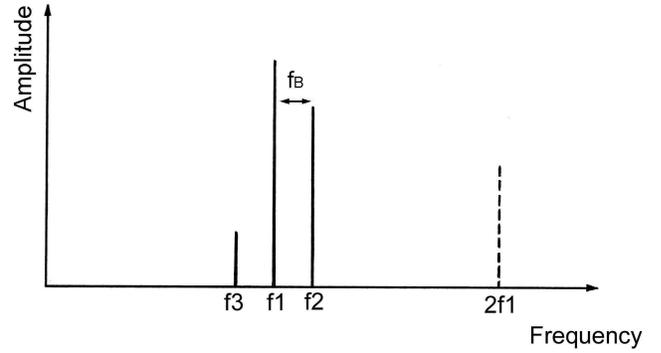} 
\caption{The triplet
  frequency $f_3$ results from the combination $2f_1~-~f_2$. We examine the departure from equidistance, $\delta f = f_2+f_3-2f_1$.}
\end{figure}

A further argument concerns the amplitude of the observed third frequency, which is either the
combination frequency itself or a resonance excitation near the combination frequency. Regrettably, there
are no one-to-one relationships between the amplitudes of the 
combination and the main frequencies involved in the combinations (Garrido \&
Rodr\'iguez 1996). Nonlinear pulsation favours the combination frequency
terms  $(if_1+jf_2)$ where both $i$ and $j$ are positive (Antonello \& Aikawa 1998).
Observational confirmation can be found for double mode cepheids (Pardo \& Poretti 1997),
and in Delta Scuti stars (Table 4 of Breger et al. 2005). The difference in combination mode amplitude
sizes even hold for a triple-mode AC And-type pulsator, V823 Cas (Jurcsik et al. 2006).

In case of resonance excitation it might be possible to see both the combination frequency and
the mode excited by resonance at nearly the same frequency. Due to the closeness of the two
frequencies, they might not be detected as two separate frequencies. Nevertheless, if both
are present, the observer sees a single frequency with very long-term amplitude modulation with
timescales of decades. Through phase locking, the resonant mode coupling may eventually result in a state of constant amplitude
(e.g., as proposed for bump cepheids, Buchler et al. 1990).

In the case of an exact frequency match we have two possibilities: the combination mode hypothesis and resonant coupling leading to a
steady limit cycle.

\section{Application to the $\delta$~Scuti variable 1 Mon}

\subsection{Available data and frequency analysis}

The $\delta$~Scuti variable 1 Mon is of special interest because it shows an almost equidistantly spaced frequency triplet
with a separation of 0.1291 cd$^{-1}$ (1.49 $\mu$Hz). It also has a low projected rotational velocity of 14 km/s (Balona et al. 2001).

An attractive interpretation of the frequency equidistance in terms of rotational splitting
of a single nonradial mode
fails, since the frequency triplet is composed of modes with different
quantum numbers, $\ell$ (Balona \& Stobie 1980, Balona et al. 2001). In particular,
they have identified the dominant 7.35 cd$^{-1}$ mode as a radial mode, which
cannot show mode splitting.  However, as Buchler et al. (1997) note, an
observed multiplet of frequencies locked through a resonance is not
necessarily associated with modes of the same degree $\ell$. 
A general condition for such coupling is given in equation (3.11) by Dziembowski (1982).

Furthermore, an explanation in terms of successive radial orders fails for several reasons,
e.g., the frequency differences are too small by more than a factor of ten compared to predictions from the mean density of the star.

Nevertheless, the equidistance or near-equidistance should not be dismissed as an accidental
agreement between three modes and provides an excellent opportunity to test the combination
mode hypothesis. A key question is whether or not the frequencies are exactly equidistant.

To examine the nature of the frequency triplet, extensive observational studies with
a long time base are needed. For the present
analysis of 1 Mon, data from the following observing seasons were used:
1970/71 and 1971/72 (Millis 1973), 1972/73  (Shobbrook \& Stobie 1974), 1976/77 and 1977/78 (Balona \& Stobie 1980),
1978/79, 1982/83 and 1984/85 (Breger \& Fedotov 2006). The less extensive photometric
data for the year 2000 (Balona et al 2001) were not used
since they are not publicly available. Hereafter, the data sets used are referred to as Mi, SS, BS and BF, respectively.

\begin{table*}
\caption[]{Multifrequency solution of 1 Mon}
\begin{center}
\begin{tabular}{rcccccc}
\hline
\noalign{\smallskip}
\multicolumn{2}{c}{Frequency} &  \multicolumn{5}{c}{$V$ amplitude (mag)}\\
cd$^{-1}$ & Name & & 1969/70 to 1971/72 & 1972/3 & 1976/77 to 1978/79 & 1982/83 to 1984/85\\
 & & Data: all & Mi & SS & BS, BF & BF\\
\noalign{\smallskip}
\hline
\noalign{\smallskip}
\multicolumn{2}{c}{Main modes}\\
\noalign{\smallskip}
   7.346153 & f$_1$ & 0.092 & 0.090 & 0.091 & 0.094 & 0.097 \\
$\pm$ 0.000002\\
   7.475269 & f$_2$ & 0.064 & 0.066 & 0.065 & 0.064 & 0.062 \\
$\pm$ 0.000005\\  
   7.217116 & f$_3$ & 0.022 & 0.021 & 0.022 & 0.021 & 0.020 \\
$\pm$ 0.000005\\
\noalign{\smallskip}
\multicolumn{2}{c}{Other frequencies}\\
\noalign{\smallskip}
   6.717240 & f$_4$ &  0.004 \\ 
   14.821422 & f$_1$ + f$_2$ &  0.017 \\ 
   14.692306 & 2f$_1$ & 0.014 \\    
   14.950538 & 2f$_2$  &     0.007 \\
   14.563268 &  f$_1$ + f$_3$  &  0.005 \\     
   22.167575 & 2f$_1$ + f$_2$  & 0.003 \\    
   22.296691 & 2f$_2$ + f$_1$  &  0.003 \\ 
   7.604385 & 2f$_2$ - f$_1$ & 0.003\\   
   22.038459 & 3f$_1$ &  0.003 \\    
   0.129116 & f$_2$ - f$_1$  &   (0.002) \\     
\noalign{\smallskip}
\hline
\end{tabular}
\end{center}
\end{table*}

For the four data sets, we have analyzed the multiple frequencies using least-squares statistical methods (Period04, see Lenz \& Breger 2005). The zero-point shifts discussed in BF were applied to the Odessa data. Thirteen frequencies were found with the required significance of amplitude signal/noise ratio greater than 4.0 (Breger et al. 1993). These
frequencies are the triplet, f$_1$ to f$_3$,  discovered before, various combinations of these three frequencies and a new mode at 6.717 cd$^{-1}$. The multifrequency analysis reveals additional peaks (e.g., at 9.544 cd$^{-1}$), which could be additional pulsation modes. However, these peaks do not reach the adopted statistical significance level and are
not listed here. Our results are shown in Table 1.

\subsection{Test of equidistance}

We can now calculate the departure from exact equidistance, $\delta$f, 
as defined by Equation\,1. This needs to be compared to the statistical
uncertainties of the computed value, which depend on the accuracy of
the three frequencies,  f$_1$, f$_2$, and f$_3$. Our uncertainty values were calculated by carrying
out extensive Monte Carlo simulations of the multifrequency sinusoidal fits. The results were checked by simple least-square
error calculations which assume white noise and uncorrelated frequencies. The results of the two
methods generally agreed to within 20\% in the sense that the more realistic Monte Carlo simulations gave larger
uncertainty values. The uncertainties are listed in Table 1.

From the four data sets covering 16 years we find\\

\begin{center}
$\delta$f = 0.000079 $\pm$ 0.000001 cd$^{-1}$.\\
\end{center}
The high accuracy of $\delta$f is not surprising if one considers
the long time base and the resulting
excellent frequency resolution. 
How much does the present result depend on the choice of data sets and our
assumptions of constant amplitudes and periods? We have repeated our analysis by optimizing
the frequencies allowing for variable amplitudes in the solution (see Table 1) or even variable amplitudes
and frequencies. In both cases, the value of  $\delta$f = 0.000079 cd$^{-1}$ found earlier was obtained.

Slightly different values are computed if fewer data and a much shorter time base are used.
If we choose only the Mi, SS and BS data (up to the 1977/78 observing season),
we find $\delta$f = 0.000109 $\pm$ 0.000004 cd$^{-1}$. This compares well to the value of
$\delta$f = 0.000115 $\pm$ 0.000008 cd$^{-1}$ reported for the same shorter data set by BS using a
different multifrequency solution.

What happens if we assume equidistance? Not surprisingly, the combination mode, 2f$_1$ - f$_2$,
misses the third peak in the power spectrum completely with the new residuals still showing f$_3$. {\bf The
solution finds f$_3$, but not at the predicted 2f$_1$-f$_2$ position.}

The results indicate that the departure from equidistance is very small but statistically
extremely significant. The departure from equidistance amounts to 0.9 nHz or 6.8s in a 3.3h period.
This also means that f$_3$ is not identical to the combination frequency (2f$_1$ - f$_2$). Since
the related combination (2f$_1$ + f$_2$) has an observed amplitude of 0.003 mag, an amplitude of
0.022 mag in the case of such an identification would have been difficult to explain. Consequently, the
excitation by resonance explanation becomes very attractive: the
combination frequency excites a nonradial mode (otherwise near stability) of almost the same frequency.

\section{Application to the RR Lyrae Blazhko star RR Lyr}
A considerable fraction of the RR Lyrae stars (20-30\% of the galactic RRab
stars - fundamental radial mode pulsators - 
and about 5\% of the RRc stars - first overtone radial mode pulsators) show a periodic amplitude
and/or phase modulation, called the Blazhko effect.  This phenomenon was
discovered a century ago, and is still not well understood.
The frequency spectra derived from externsive light and radial velocity curves of RR 
Lyrae Blazhko stars exhibit either a doublet structure or an equally-spaced triplet
structure around the main pulsation frequency $f_1$ as well as its harmonics $kf_1$.
The small frequency separation in the triplets/doublets corresponds to the 
Blazhko or modulation frequency $f_B$ (see Kov\'acs 2001; Alcock et al. 2000,
2003).  Since the main peak corresponds to a radial mode, rotational splitting
can be ruled out.
It is an important question as to why the multiplet structures occur with equal 
spacings at the harmonics just as at the main frequency.

\subsection{Resonance models and combination mode hypothesis for RR Lyrae stars}

The currently most cited models for explaining the Blazhko effect involve
either a magnetic field or a resonance effect.  In both types of models
nonradial oscillation modes play a crucial role.  The present resonance models 
suggest
that the Blazhko effect is caused by a direct resonance between the main
radial mode and a nonradial mode, most likely of a low degree $\ell$. 
Van Hoolst et al. (1998) showed that RR Lyrae stars have a very dense
nonradial frequency spectrum, and that nonradial modes in the vicinity of the
radial mode have a lower moment of inertia and hence are more prone to
instability. Nowakowski \& Dziembowski
(2001) investigated resonant excitation of nonradial modes in RR Lyrae stars
and predicted significant amplitude (and phase) modulation in the case of
excitation of a nonradial mode of low degree close to the radial 
mode. However, the strong nonlinear damping of nonradial modes at the
amplitudes observed poses a major problem
for such models (Nowakowsi \& Dziembowski 2003). Dziembowski \&
Mizerski (2004) propose a scenario for the Blazhko effect, involving energy 
transfer from the radial to nonradial modes, but avoiding the problems of
finite amplitude development uncovered by Nowakowski \& Dziembowski (2003). 

For RR Lyrae stars a simple combination mode hypothesis was proposed by
Borkowski (1981), who interpreted the numerous components in the light
variation of AR Her as a nonlinear superposition of the main radial frequency $f_1$, and
a mode at $2f_1+f_B$.  The additional mode was at first identified
by Borkowski (1981) as the second or third radial overtone of the star.
Later on, model calculations yielded the radial overtone modes at different
frequencies and hence the additional mode must be a nonradial one (Kov\'acs
1995).
Recently, Kolenberg et al. (2006, hereafter K06) proposed a similar hypothesis to
interpret the light curve variations of RR Lyrae as a combination of the
main radial mode with a frequency $f_1$ and a nonradial mode at $f_N=f_1+f_B$.
These are the same frequencies as were considered by Bal\'asz \& Detre (1943) to
be at the basis of the amplitude modulation.

Alcock et al. (2000, 2003) examined the equidistance of triplet 
frequencies on a sample of RRc and RRab Blazhko stars found 
in the MACHO database.  They defined the departure from equidistance as $f_{+}
+ f_{-} - 2 f_0$, where $f_0$, $f_{-}$ and $f_{+}$ denote the main frequency
(here $f_1$), the lower and the higher frequency side peak, respectively.
For the RRc stars they found the deviations from 
equidistance to be insignificant (Alcock et al. 2000). 
For the RRab stars, the quantity  $\delta f$ was, with
few exceptions, within $\pm$0.0002 c/d, half of the characteristic line width
(Alcock et al. 2003).
However, an analysis of artificial time series led to the conclusion
that the observed small deviations from
equidistant spacing cannot be accounted for by purely noise-induced frequency
shifts.
The authors state that the nonequidistant spacing has various origins, 
including cases in which neither the noise level nor peculiar additional 
features are suspected to be responsible for the deviation.  
Hence it is recommended to study the cause of non-equidistant frequency 
spacing on a star-by-star basis.  

\subsection{Decomposing the RR Lyrae frequency spectrum}

We now examine the case of RR Lyr, the
brightest Blazhko star, for which we used the photometric
data published by K06.  These data were
obtained over a 421-day interval in 2003-2004.  The data from only 2004 cover
a 194-day interval, i.e., about 5 Blazhko cycles.
RR Lyr is known to have a radial pulsation period $f_1=1.7642$ c/d.  Its Blazhko
or modulation period used to be about 40.8 days, but recent data show that it
has decreased to about 39 days (K06).

For RR Lyrae Blazhko stars, usually a fit assuming intrinsically
  equidistant triplet frequencies is proposed, according to the following
  model:
\begin{equation}
\renewcommand{\arraystretch}{1.2}\begin{array}{l}
f(t) = A_0 + \sum_{k=1}^n [A_k \sin (2 \pi kf_1 t + \phi_k)\\
+ A_{k+} \sin (2 \pi (kf_1 + f_B)t + \phi_{k+})\\
+ A_{k-} \sin (2 \pi (kf_1 - f_B)t + \phi_{k-})]\\
+ B_0 \sin (2 \pi (f_B t + \phi_B)),
\end{array}
\end{equation}
where $f_1$ is the main frequency and $f_B$ the Blazhko frequency. 
The results of the fit according to Equation\,2 were published by 
K06.  They represent the common way of decomposing the frequency spectrum of a
Blazhko star.

We now examine the recent RR Lyr data set published by K06 in terms of the 
equidistance of the triplet frequencies.  We first consider the first order
triplet, i.e., the triplet at the main
frequency.  The triplets at higher orders are considered in subsection 4.4. 
As Alcock et al. (2003) concluded from their extensive study on the
LMC fundamental mode RR Lyrae Blazhko stars, the higher (right-hand side) frequencies in the 
triplets have higher amplitudes in about 75\% of the cases.  RR Lyr also falls among
the stars with  right-side peaks higher than the left-side peaks.

\subsection{Equidistance at the first order triplet}

Instead of assuming a least-squares fit according to Equation\,2, we
allow triplet structures which may not be exactly equidistant, leaving
  free the exact value of the side peak frequencies. Hence we
construct a fit as follows:
\begin{equation}
\renewcommand{\arraystretch}{1.2}\begin{array}{l}
f(t) = A_0 + \sum_{k=1}^n [A_k \sin (2 \pi kf_1 t + \phi_k)\\
+ A_{k+} \sin (2 \pi (f_{k+})t + \phi_{k+})\\
+ A_{k-} \sin (2 \pi (f_{k-})t + \phi_{k-})]\\
+ B_0 \sin (2 \pi (f_B t + \phi_B)).
\end{array}
\end{equation}

As for 1 Mon, the noise-induced frequency shifts were determined by means
of Monte Carlo simulations with extensive numbers of iterations to obtain
stability of the calculated values of the uncertainty.
The departure from equidistance in the first order triplet is $\delta
f=0.00072 \pm 0.00018$ c/d.  This result calls for caution in the
standard fitting of Blazhko stars with equidistant triplet structures.

\subsection{Extension to higher order triplets}

We now consider the departure from equidistance, 
$\delta f$, for the higher order triplets up to 
the fourth order, i.e., the components having significant 
amplitudes. Table\,2 gives the departures from equidistance calculated in the standard way
($\delta$f = f$_2$+f$_3$-2f$_1$), column 4 in Table\,2.

However, in the combination hypothesis for the higher order side peaks, it 
would be more appropriate to
calculate the departure from the expected position of the combination
frequency, $\delta f_C$, rather than checking the departure from
equidistance.  For the first order triplet both terms yield the same
value, but for the higher order triplets the values are different.

According
to the combination hypothesis, the side peaks result
from a sum (right side peaks) or difference (left side peaks) combination of
the nonradial component $f_N$ with $kf_1$ with $k=1,2,...$ (up to the highest
significant harmonic order).  Hence, the departure from the expected position
of the combination frequency should be calculated for each
of the side peaks (left and right) separately. 
For example, in the combination hypothesis $f_{2-}=2f_1-f_B$, the
left side peak at twice the main frequency is equal to $3f_1-f_N$.  The frequency is observed at
$f_{2-}= 3.50260 \pm 0.00012$ c/d.  The frequency $3f_1-f_N$ is expected to occur
at $3.502442 \pm 0.000028$ c/d.  Hence the departure from equidistance $\delta
f_C$ is equal to
$-0.00016 \pm 0.00012$ c/d. The departure from equidistance $\delta f$ yields 
a different value, $0.00015 \pm 0.00013$ c/d. 
A similar reasoning holds for the right side peaks in the triplets, and for
all of the higher order triplet components.

The combination hypothesis may be a plausible explanation for the
triplet structures in RR Lyrae stars. Hence, it makes sense to consider an
additional parameter for the higher order triplets, in addition to the 
previously defined
one which is valid for the triplet around the main frequency, and which we
apply in this paper to different types of pulsating stars. 

As can be seen from Table\, 2, the departure from the expected position
  of the combination frequency for the higher order triplet structures 
is significant 
for some components, while for others it is not.  The uncertainties given are
derived from the uncertainties on the observed higher order side peak 
frequencies and the (first order) frequencies involved in the
combination. They represent 1-$\sigma$ uncertainties.  

In general, we can conclude that the triplets in RR Lyrae Blazhko stars are
not necessarily exactly equidistant, in accordance with the findings by Alcock
et al. (2003).  This should be considered in future
studies of Blazhko stars.  The departure from equidistance and/or from the
expected position of the combination frequency may provide a test for the
combination mode hypothesis to explain the typical frequency spectrum of a
Blazhko star.

\begin{table*}
\caption{Triplet freqencies, their uncertainties and the derived deviations based on the 2004 RR~Lyr data
  published by Kolenberg et al. (2006). The departure from
  equidistance, $\delta f$, is given for the first four triplet orders. 
The last two columns give the departure from the expected position of the
  combination frequency
  for each of the side peak frequencies. } 
\label{table:1}      
\centering                          

\begin{tabular}{c c c c c c c c}        
\hline
\hline               
 \multicolumn{2}{c}{Triplet order} & \multicolumn{3}{c}{Triplet frequencies} & \multicolumn{3}{c}{Deviation 
from:} \\
&& Central & Left & Right & Equidistancy &  \multicolumn{2}{c}{Expected combination}\\
& $k$ & $f_1$ & $f_2$ & $f_3$ &  $\delta f$ &Left&Right   \\
\hline                        
First order & 1  &	1.7641849 &1.737533 & 1.790113 & 0.00072 & 0.00072& - \\
& & $\pm$0.0000022&$\pm$0.00018&$\pm$0.000028&$\pm$ 0.00018 &$\pm$ 0.00018& -\\
\hline
Higher order & 2  &	3.5283698 &3.50260& 3.553953& 0.00015& -0.00016 & 0.00034  	 \\
& & $\pm$0.0000044&$\pm$0.00012&$\pm$0.000049&$\pm$ 0.00013 & $\pm$0.00012 &
$\pm$ 0.000049\\
& 3  &	5.2925547 & 5.266189 & 5.318550	& -0.00037 & 0.00043 & -0.000067 \\
& & $\pm$0.0000066&$\pm$0.00010& $\pm$0.000043&$\pm$0.00011  & $\pm$0.00010 & $\pm$0.000051 \\
& 4  &	7.0567396 &7.030806 &7.082288	&-0.00038  & 0.0000053 & 0.00038 \\
& & $\pm$0.0000088&$\pm$0.00016&$\pm$0.000049 & $\pm$0.00017& $\pm$0.00016 & $\pm$0.000057  \\
\hline                                \end{tabular}
\end{table*}

\section{The cepheid V743 Lyr}

V743 Lyr (HR 7308) is probably the best-studied cepheid with a strong Blazhko Effect. This 1.49d cepheid has
a modulation period near 1200d. Breger (1981) as well as Burki, Mayor \& Benz (1982) have shown that
a simple model based on beating between two closely spaced frequencies cannot explain the observed phase and
amplitude variability. Furthermore, the long modulation period rules out rotational splitting of nonradial modes as well.
An equidistant or near-equidistant frequency triplet has been reported for V743 Lyr: 
Koen (2001) analyzed the available Hipparcos photometry and proposed
that the amplitude variations of V743 Lyr could be described by symmetrical frequency triplets separated
by 0.0011 c/d. 

We have collected the available photometric data from 1966 to 1983 (Percy \& Evans 1980,
Breger 1981, Henrikssen 1983, Burki et al. 1986, Breger 2006). We also included the excellent radial-velocity data
by Burki, Mayor \& Benz (1982). The periodicities in the data were analyzed with the
PERIOD04 statistical package as in Section 2.

The multifrequency analysis confirmed the existence of the frequency triplet (0.67075, 0.67157, and 0.66988 cd$^{-1}$).
The frequency triplet misses equidistance by
0.00005 $\pm$ 0.00001 cd$^{-1}$. Because of the small formal error the result appears to be
statistically significant. However, the solution may not be less certain for the following reasons:
(i) The fits are not good: the standard deviation is 0.023 mag per single measurement.
(ii) Even after the inclusion of the radial-velocity data, the spectral window indicates that
the frequencies and their harmonics are not statistically independent of each other.
The unfavorable main pulsation period of 1.49d represents a severe problem.
(iii) Analysis of the residuals shows additional periodicities near the main frequencies.

We conclude that the data for the best-studied Blazhko cepheid are insufficient to test the combination mode hypothesis.
	
\section{Conclusion}

We have examined the phenomenon of (nearly) equidistant frequency triplets in
a $\delta$~Scuti star, an RR~Lyrae star, and a cepheid variable and tested whether the smallest amplitude component could be a combination of the
other two modes. For the $\delta$~Scuti variable 1 Mon, a departure from equidistance  of only
0.000079 $\pm$ 0.000001 cd$^{-1}$ (or 0.91 $\pm$ 0.01 nHz). Here the Combination Mode Hypothesis with a mode
excited by resonance is the most probable explanation. For the star RR~Lyr, the hypothesis of resonance
through a combination of modes was also found to be plausible. Finally, the results for the best-studied cepheid with a Blazhko period (V743 Lyr)
were inconclusive because of an unfavorable period of 1.49d and insufficient data.

\section*{Acknowledgments} Part of the investigation has been supported by the
Austrian Fonds zur F\"{o}rderung der wissenschaftlichen Forschung.

\end{document}